\documentclass[12pt]{article}
\usepackage{axodraw}

\catcode`@=12
\begin{document}
\thispagestyle{empty}
\begin{flushright} 
UCRHEP-T366\\ 
November 2003\
\end{flushright}
\vspace{0.5in}
\begin{center}
{\LARGE	\bf The Neutrino Mass Matrix -- \\From $A_4$ to $Z_3$\\}
\vspace{1.5in}
{\bf Ernest Ma\\}
\vspace{0.2in}
{\sl Physics Department, University of California, Riverside, 
California 92521\\}
\vspace{2.0in}
\end{center}
\begin{abstract}\
With the recent experimental advance in our precise knowledge of the 
neutrino oscillation parameters, the correct form of the $3 \times 3$ 
neutrino mass matrix is now approximately known.  I discuss how this may 
be obtained from symmetry principles, using as examples the finite groups 
$A_4$ and $Z_3$, in two complete theories of leptons (and quarks).
\end{abstract}

---------

Preprint version of talk at the 9th Adriatic Meeting in Dubrovnik (2003).

\newpage
\baselineskip 24pt

\section{Introduction}

After the new experimental results of KamLAND \cite{kamland} on top of 
those of SNO \cite{sno} and SuperKamiokande \cite{atm}, etc. \cite{reactor}, 
we now have very good knowledge of 5 parameters:
\begin{eqnarray}
&& \Delta m^2_{atm} \simeq 2.5 \times 10^{-3}~{\rm eV}^2, \\ 
&& \Delta m^2_{sol} \simeq 6.9 \times 10^{-5}~{\rm eV}^2, \\ 
&& \sin^2 2 \theta_{atm} \simeq 1, \\ 
&& \tan^2 \theta_{sol} \simeq 0.46, \\ 
&& |U_{e3}| < 0.16.
\end{eqnarray}
The last 3 numbers tell us that the neutrino mixing matrix is rather 
well-known, and to a very good first approximation, it is given by
\begin{equation}
\pmatrix {\nu_e \cr \nu_\mu \cr \nu_\tau} = \pmatrix {c & -s & 0 \cr 
s/\sqrt 2 & c/\sqrt 2 & -1/\sqrt 2 \cr s/\sqrt 2 & c/\sqrt 2 & 1/\sqrt 2} 
\pmatrix {\nu_1 \cr \nu_2 \cr \nu_3},
\end{equation}
where $\sin^2 2 \theta_{atm} =1$ and $U_{e3} = 0$ have been assumed, with 
$s \equiv \sin \theta_{sol}$, $c \equiv \cos \theta_{sol}$.

\section{Approximate Generic Form of the Neutrino Mass Matrix}

Assuming three Majorana neutrino mass eigenstates with real eigenvalues 
$m_{1,2,3}$, the neutrino mass matrix in the basis $(\nu_e,\nu_\mu,\nu_\tau)$ 
is then of the form \cite{ma02}
\begin{equation}
{\cal M}_\nu = \pmatrix {a+2b+2c & d & d \cr d & b & a+b \cr d & a+b & b}.
\end{equation}
Note that ${\cal M}_\nu$ is invariant under the discrete $Z_2$ symmetry: 
$\nu_e \to \nu_e$, $\nu_\mu \leftrightarrow \nu_\tau$. 
Depending on the relative magnitudes of the 4 parameters $a,b,c,d$, this 
matrix has 7 possible limits:  3 have the normal hierarchy, 2 have the 
inverted hierarchy, and 2 have 3 nearly degenerate masses.

In neutrinoless double beta decay, the effective mass is $m_0 = |a+2b+2c|$. 
In the 2 cases of inverted hierarchy, we have
\begin{eqnarray}
&& m_0 \simeq \sqrt {\Delta m^2_{atm}} \simeq 0.05~{\rm eV}, \\ 
&& m_0 \simeq \cos 2 \theta_{sol} \sqrt {\Delta m^2_{atm}},
\end{eqnarray}
respectively for $m_1/m_2 = \pm 1$, i.e. for their relative $CP$ being even 
or odd.  In the 2 degenerate cases,
\begin{eqnarray}
&& m_0 \simeq |m_{1,2,3}|, \\ 
&& m_0 \simeq \cos 2 \theta_{sol} |m_{1,2,3}|.
\end{eqnarray}

With ${\cal M}_\nu$ of Eq.~(7), $U_{e3}$ is zero necessarily, in which case 
there can be no $CP$ violation in neutrino oscillations.  However, suppose 
we consider instead \cite{ma02,gl03}
\begin{equation}
{\cal M}_\nu = \pmatrix {a+2b+2c & d & d^* \cr d & b & a+b \cr d^* & a+b & b},
\end{equation}
where $d$ is now complex, then the $Z_2$ symmetry of Eq.~(7) is broken and 
$U_{e3}$ becomes nonzero.  In fact, it is proportional to $i Im d$, thus 
predicting maximal $CP$ violation in neutrino oscillations.

\section{Nearly Degenerate Majorana Neutrino Masses}

Suppose that at some high energy scale, the charged lepton mass matrix and 
the Majorana neutrino mass matrix are such that after diagonalizing the 
former, i.e.
\begin{equation}
{\cal M}_l = \pmatrix {m_e & 0 & 0 \cr 0 & m_\mu & 0 \cr 0 & 0 & m_\tau},
\end{equation}
the latter is of the form
\begin{equation}
{\cal M}_\nu = \pmatrix {m_0 & 0 & 0 \cr 0 & 0 & m_0 \cr 0 & m_0 & 0}.
\end{equation}
From the high scale to the electroweak scale, one-loop radiative corrections 
will change ${\cal M}_\nu$ as follows:
\begin{equation}
({\cal M}_\nu)_{ij} \to ({\cal M}_\nu)_{ij} + R_{ik} ({\cal M}_\nu)_{kj} 
+ ({\cal M}_\nu)_{ik} R^T_{kj},
\end{equation}
where the radiative correction matrix is assumed to be of the most general 
form, i.e.
\begin{equation}
R = \pmatrix {r_{ee} & r_{e\mu} & r_{e\tau} \cr r_{e\mu}^* & r_{\mu\mu} & 
r_{\mu\tau} \cr r_{e\tau}^* & r_{\mu\tau}^* & r_{\tau\tau}}.
\end{equation}
Thus the observed neutrino mass matrix is given by
\begin{equation}
{\cal M}_\nu = m_0 \pmatrix {1+2r_{ee} & r_{e\tau} + r_{e\mu}^* & r_{e\mu} + 
r_{e\tau}^* \cr r_{e\mu}^* + r_{e\tau} & 2r_{\mu\tau} & 1+r_{\mu\mu}+
r_{\tau\tau} \cr r_{e\tau}^* + r_{e\mu} & 1+r_{\mu\mu}+r_{\tau\tau} & 
2r_{\mu\tau}^*}.
\end{equation}
Let us rephase $\nu_\mu$ and $\nu_\tau$ to make $r_{\mu \tau}$ real, then 
the above ${\cal M}_\nu$ is exactly in the form of Eq.~(12), with of course 
$a$ as the dominant term.  In other words, we have obtained a desirable 
description of all present data on neutrino oscillations including $CP$ 
violation, starting from almost nothing.

\section{Plato's Fire}

The successful derivation of Eq.~(17) depends on having Eqs.~(13) and (14). 
To be sensible theoretically, they should be maintained by a symmetry. 
At first sight, it appears impossible that there can be a symmetry which 
allows them to coexist.  The solution turns out to be the non-Abelian 
discrete symmetry $A_4$ \cite{mrmm,bmv}. What is $A_4$ and why is it special?

Around the year 390 BCE, the Greek mathematician Theaetetus proved that there 
are five and only five perfect geometric solids.  The Greeks already knew 
that there are four basic elements: fire, air, water, and earth.  Plato 
could not resist matching them to the five perfect geometric solids and 
for that to work, he invented the fifth element, i.e. quintessence, which 
is supposed to hold the cosmos together.  His assignments are shown in 
Table 1.

\begin{table}[htb]
\caption{Properties of Perfect Geometric Solids}
\begin{center}
\begin{tabular}{|c|c|c|c|c|}
\hline 
solid & faces & vertices & Plato & Group \\ 
\hline
tetrahedron & 4 & 4 & fire & $A_4$ \\ 
octahedron & 8 & 6 & air & $S_4$ \\ 
icosahedron & 20 & 12 & water & $A_5$ \\ 
hexahedron & 6 & 8 & earth & $S_4$ \\ 
dodecahedron & 12 & 20 & ? & $A_5$ \\ 
\hline
\end{tabular}
\end{center}
\end{table}

The group theory of these solids was established in the early 19th century. 
Since a cube (hexahedron) can be imbedded perfectly inside an octahedron 
and the latter inside the former, they have the same symmetry group.  
The same holds for the icosahedron and dodecahedron.  The tetrahedron 
(Plato's ``fire'') is special because it is self-dual.  It has the symmetry 
group $A_4$, i.e. the finite group of the even permutation of 4 objects. 
The reason that it is special for the neutrino mass matrix is because it 
has three inequivalent one-dimensional irreducible representations and one 
three-dimensional irreducible representation exactly.  Its character table 
is given below.

\begin{table}[htb]
\caption{Character Table of $A_4$}
\begin{center}
\begin{tabular}{|c|c|c|c|c|c|c|}
\hline
class & n & h & $\chi_1$ & $\chi_2$ & $\chi_3$ & $\chi_4$ \\ 
\hline
$C_1$ & 1 & 1 & 1 & 1 & 1 & 3 \\ 
$C_2$ & 4 & 3 & 1 & $\omega$ & $\omega^2$ & 0 \\ 
$C_3$ & 4 & 3 & 1 & $\omega^2$ & $\omega$ & 0 \\ 
$C_4$ & 3 & 2 & 1 & 1 & 1 & $-1$ \\ 
\hline
\end{tabular}
\end{center}
\end{table}

In the above, $n$ is the number of elements, $h$ is the order of each element, 
and
\begin{equation}
\omega = e^{2 \pi i/3}
\end{equation}
is the cube root of unity.  The group multiplication rule is
\begin{equation}
\underline {3} \times \underline {3} = \underline {1} + \underline {1}' + 
\underline {1}'' + \underline {3} + \underline {3}.
\end{equation}

\section{Details of the $A_4$ Model}

The fact that $A_4$ has three inequivalent 
one-dimensional representations \underline {1}, \underline {1}$'$, 
\underline {1}$''$, and one three-dimensional reprsentation \underline {3}, 
with the decomposition given by Eq.~(19) leads naturally to the following 
assignments of quarks and leptons:
\begin{eqnarray}
&& (u_i,d_i)_L, ~~ (\nu_i,e_i)_L \sim \underline {3}, \\ 
&& u_{1R}, ~~ d_{1R}, ~~ e_{1R} \sim \underline {1}, \\ 
&& u_{2R}, ~~ d_{2R}, ~~ e_{2R} \sim \underline {1}', \\ 
&& u_{3R}, ~~ d_{3R}, ~~ e_{3R} \sim \underline {1}''.
\end{eqnarray}
Heavy fermion singlets are then added:
\begin{equation}
U_{iL(R)}, ~~ D_{iL(R)}, ~~ E_{iL(R)}, ~~ N_{iR} \sim \underline {3},
\end{equation}
together with the usual Higgs doublet and new heavy singlets:
\begin{equation}
(\phi^+,\phi^0) \sim \underline {1}, ~~~~ \chi^0_i \sim \underline {3}.
\end{equation}
With this structure, charged leptons acquire an effective Yukawa coupling 
matrix $\bar e_{iL} e_{jR} \phi^0$ which has 3 arbitrary eigenvalues 
(because of the 3 independent couplings to the 3 inequivalent one-dimensional 
representations) and for the case of equal vacuum expectation values of 
$\chi_i$, i.e.
\begin{equation}
\langle \chi_1 \rangle = \langle \chi_2 \rangle = \langle \chi_3 \rangle = u,
\end{equation}
which occurs naturally in the supersymmetric version of this model \cite{bmv}, 
the unitary transformation $U_L$ which diagonalizes ${\cal M}_l$ is given by
\begin{equation}
U_L = {1 \over \sqrt 3} \pmatrix {1 & 1 & 1 \cr 1 & \omega & \omega^2 \cr 
1 & \omega^2 & \omega}.
\end{equation}
This implies that the effective neutrino 
mass operator, i.e. $\nu_i \nu_j \phi^0 \phi^0$, is proportional to
\begin{equation}
U_L^T U_L = \pmatrix {1 & 0 & 0 \cr 0 & 0 & 1 \cr 0 & 1 & 0},
\end{equation}
exactly as desired.

\section{New Flavor-Changing Radiative Mechanism}

The original $A_4$ model \cite{mrmm} was conceived to be a symmetry at the 
electroweak scale, in which case the splitting of the neutrino mass 
degeneracy is put in by hand and any mixing matrix is possible.  Subsequently, 
it was proposed \cite{bmv} as a symmetry at a high scale, in which case the 
mixing matrix is determined completely by flavor-changing radiative 
corrections and the only possible result happens to be Eq.~(17).  This is a 
remarkable convergence in that Eq.~(17) is in the form of Eq.~(12), i.e. the 
phenomenologically preferred neutrino mixing matrix based on the most recent 
data from neutrino oscillations.

We should now consider the new physics responsible for the $r_{ij}$'s of 
Eq.~(16).  Previously \cite{bmv}, arbitrary soft supersymmetry breaking in 
the scalar sector was invoked.  It is certainly a phenomenologically viable 
scenario, but lacks theoretical motivation and is somewhat complicated.  Here 
a new and much simpler mechanism is proposed \cite{plato}, using a triplet 
of charged scalars under $A_4$, i.e. $\eta^+_i \sim \underline {3}$.  Their 
relevant contributions to the Lagrangian of this model is then
\begin{equation}
{\cal L} = f \epsilon_{ijk} (\nu_i e_j - e_i \nu_j) \eta^+_k + m_{ij}^2 
\eta^+_i \eta^-_j.
\end{equation}
Whereas the first term is invariant under $A_4$ as it should be, the second 
term is a soft term which is allowed to break $A_4$, from which the 
flavor-changing radiative corrections will be calculated.

Let
\begin{equation}
\pmatrix {\eta_e \cr \eta_\mu \cr \eta_\tau} = \pmatrix {U_{e1} & U_{e2} & 
U_{e3} \cr U_{\mu 1} & U_{\mu 2} & U_{\mu 3} \cr U_{\tau 1} & U_{\tau 2} & 
U_{\tau 3}} \pmatrix {\eta_1 \cr \eta_2 \cr \eta_3},
\end{equation}
where $\eta_{1,2,3}$ are mass eigenstates with masses $m_{1,2,3}$.  The 
resulting radiative corrections are given by
\begin{equation}
r_{\alpha \beta} = -{f^2 \over 8 \pi^2} \sum_{i=1}^3 U^*_{\alpha i} 
U_{\beta i} \ln m_i^2.
\end{equation}
To the extent that $r_{\mu \tau}$ should not be larger than about $10^{-2}$, 
the common mass $m_0$ of the three degenerate neutrinos should not be less 
than about 0.2 eV in this model.  This is consistent with the recent 
WMAP upper bound \cite{wmap} of 0.23 eV and the range 0.11 to 0.56 eV 
indicated by neutrinoless double beta decay \cite{klapdor}.

\section{Models based on $S_3$ and $D_4$}

Two other examples of the application of non-Abelian discrete symmetries to 
the neutrino mass matrix have recently been proposed.  One \cite{kmmrj} is 
based on the symmetry group of the equilateral triangle $S_3$, which has 
6 elements and the irreducible representations \underline {1}, 
\underline {1}$'$, and \underline {2}.  The 3 families of leptons as well 
as 3 Higgs doublets transform as \underline {1} + \underline {2} under $S_3$. 
An additional $Z_2$ is introduced where $\nu_R$(\underline {1}) and 
$H$(\underline {2}) are odd, while all other fields are even.  After a 
detailed analysis, the mixing matrix of Eq.~(6) is obtained with $U_{e3} 
\simeq -3.4 \times 10^{-3}$ and $0.4 < \tan \theta_{sol} < 0.8$.  The 
neutrino masses are predicted to have an inverted hierarchy satisfying 
Eq.~(8).

Another example \cite{gld4} is based on the symmetry group of the square 
$D_4$, which has 8 elements and the irreducible representations 
\underline {1}$^{++}$, \underline {1}$^{+-}$, \underline {1}$^{-+}$, 
\underline {1}$^{--}$, and \underline {2}.  The 3 families of leptons 
transform as \underline {1}$^{++}$ + \underline {2}.  The Higgs sector has 
3 doublets with $\phi_3 \sim$ \underline {1}$^{+-}$ and 2 singlets $\chi \sim$ 
\underline {2}.  Under an extra $Z_2$, $\nu_R$, $e_R$, $\phi_1$ are odd, while 
all other fields are even, including $\phi_2$.  This results in the neutrino 
mass matrix of Eq.~(7) with an additional constraint, i.e. $m_1 < m_2 < m_3$ 
such that the $m_0$ of neutrinoless double beta decay is equal to $m_1 m_2
/m_3$.

\section{Form Invariance of the Neutrino Mass Matrix}

Consider a specific $3 \times 3$ unitary matrix $U$ and impose the condition
\cite{ma03}
\begin{equation}
U {\cal M}_\nu U^T = {\cal M}_\nu
\end{equation}
on the neutrino mass matrix ${\cal M}_\nu$ in the $(\nu_e,\nu_\mu,\nu_\tau)$ 
basis.  Iteration of the above yields
\begin{equation}
U^n {\cal M}_\nu (U^T)^n = {\cal M}_\nu.
\end{equation}
Therefore, unless $U^{\bar n} = 1$ for some finite $\bar n$, the only solution 
for ${\cal M}_\nu$ would be a multiple of the identity matrix.  Take for 
example $\bar n = 2$, then the choice
\begin{equation}
U = \pmatrix {1 & 0 & 0 \cr 0 & 0 & 1 \cr 0 & 1 & 0}
\end{equation}
leads to Eq.~(7).  In other words, the present neutrino oscillation data may 
be understood as a manifestation of the discrete symmetry $\nu_e \to \nu_e$ 
and $\nu_\mu \leftrightarrow \nu_\tau$.

Suppose instead that $\bar n = 4$, with $U^2$ given by Eq.~(34), then one 
possible solution for its square root is
\begin{equation}
U_1 = \pmatrix {1 & 0 & 0 \cr 0 & (1-i)/\sqrt 2 & (1+i)/\sqrt 2 \cr 0 & 
(1+i)/\sqrt 2 & (1-i)/\sqrt 2},
\end{equation}
which leads to
\begin{equation}
{\cal M}_1 = \pmatrix {2b+2c & d & d \cr d & b & b \cr d & b & b},
\end{equation}
i.e. the 4 parameters of Eq.~(7) have been reduced to 3 by setting $a=0$.

Another solution is
\begin{equation}
U_2 = {1 \over \sqrt 3} \pmatrix {1 & 1 & 1 \cr 1 & \omega & \omega^2 \cr 
1 & \omega^2 & \omega},
\end{equation}
which leads to
\begin{equation}
{\cal M}_2 = \pmatrix {2b+2d & d & d \cr d & b & b \cr d & b & b},
\end{equation}
i.e. ${\cal M}_1$ has been reduced by setting $c=d$.  The 3 mass 
eigenvalues are then $m_{1,2} = 2b \mp \sqrt 2 d$ and $m_3=0$, i.e. 
an inverted hierarchy, with $\tan^2 \theta_{sol}$ predicted to be 
$2 - \sqrt 3 = 0.27$, as compared to the allowed range \cite{msv} 
0.29 to 0.86 from fitting all present data.

\section{New $Z_3$ Model of Neutrino Masses}

Very recently, two new complete models of lepton 
masses have been obtained, one based on $Z_4$ \cite{mr03} and the other 
on $Z_3$ \cite{maz303}.  The former does not fix the solar mixing angle, 
whereas the latter predicts $\tan^2 \theta_{sol} = 0.5$.  Here I will discuss 
only the $Z_3$ case.  Let ${\cal M}_\nu$ be given by
\begin{equation}
{\cal M}_\nu = {\cal M}_A + {\cal M}_B + {\cal M}_C,
\end{equation}
where
\begin{equation}
{\cal M}_A = A \pmatrix {1 & 0 & 0 \cr 0 & 1 & 0 \cr 0 & 0 & 1}, ~~ 
{\cal M}_B = B \pmatrix {-1 & 0 & 0 \cr 0 & 0 & -1 \cr 0 & -1 & 0}, 
~~ {\cal M}_C = C \pmatrix {1 & 1 & 1 \cr 1 & 1 & 1 \cr 1 & 1 & 1}.
\end{equation}
Since the invariance of ${\cal M}_A$ requires only $U_A U_A^T = 1$, $U_A$ can 
be any orthogonal matrix.  As for ${\cal M}_B$ and ${\cal M}_C$, they are 
both invariant under the $Z_2$ transformation of Eq.~(34) 
and each is invariant under a $Z_3$ transformation, i.e. 
$U_B^3 = 1$ and $U_C^3 = 1$, but $U_B \neq U_C$.  Specifically,
\begin{equation}
U_B = \pmatrix {-1/2 & -\sqrt{3/8} & -\sqrt{3/8} \cr \sqrt{3/8} & 1/4 & 
-3/4 \cr \sqrt{3/8} & -3/4 & 1/4}, ~~~  
U_C = \pmatrix {0 & 1 & 0 \cr 0 & 0 & 1 \cr 1 & 0 & 0}.
\end{equation}
Note that $U_B$ commutes with $U_2$, but $U_C$ does not. If $U_C$ is combined 
with $U_2$, then the non-Abelian discrete symmetry $S_3$ is generated.

First consider $C=0$.  Then ${\cal M}_\nu = {\cal M}_A + {\cal M}_B$ is the 
most general solution of
\begin{equation}
U_B {\cal M}_\nu U_B^T = {\cal M}_\nu,
\end{equation}
and the eigenvectors of ${\cal M}_\nu$ are $\nu_e$, $(\nu_\mu + \nu_\tau)/
\sqrt 2$, and $(\nu_\mu - \nu_\tau)/\sqrt 2$ with eigenvalues $A-B$, $A-B$, 
and $A+B$ respectively.  This explains atmospheric neutrino oscillations 
with $\sin^2 2 \theta_{atm} = 1$ and
\begin{equation}
(\Delta m^2)_{atm} = (A+B)^2 - (A-B)^2 = 4BA.
\end{equation}

Now consider $C \neq 0$.  Then in the basis spanned by $\nu_e$, $(\nu_\mu + 
\nu_\tau)/\sqrt 2$, and $(\nu_\mu - \nu_\tau)/\sqrt 2$,
\begin{equation}
{\cal M}_\nu = \pmatrix {A-B+C & \sqrt 2 C & 0 \cr \sqrt 2 C & A-B+2C & 0 \cr 
0 & 0 & A+B}.
\end{equation}
The eigenvectors and eigenvalues become
\begin{eqnarray}
\nu_1 = {1 \over \sqrt 6} (2 \nu_e - \nu_\mu - \nu_\tau), && m_1 = A-B, \\ 
\nu_2 = {1 \over \sqrt 3} (\nu_e + \nu_\mu + \nu_\tau), && m_2 = A-B+3C, \\ 
\nu_3 = {1 \over \sqrt 2} (\nu_\mu - \nu_\tau), && m_3 = A+B.
\end{eqnarray}
This explains solar neutrino oscillations as well with $\tan^2 \theta_{sol} = 
1/2$ and
\begin{equation}
(\Delta m^2)_{sol} = (A-B+3C)^2 - (A-B)^2 = 3C(2A-2B+3C).
\end{equation}
Whereas the mixing angles are fixed, the proposed ${\cal M}_\nu$ has the 
flexibility to accommodate the three patterns of neutrino masses often 
mentioned, i.e.

(I) the hierarchical solution, e.g. $B=A$ and $C << A$;

(II) the inverted hierarchical solution, e.g. $B = -A$ and $C << A$; 

(III) the degenerate solution, e.g. $C << B << A$.

\noindent In all cases, $C$ must be small. Therefore ${\cal M}_\nu$ of 
Eq.~(39) satisfies Eq.~(42) to a very good approximation, and $Z_2 \times 
Z_3$ as generated by $U_2$ and $U_B$ should be taken as the underlying 
symmetry of this model.

Since ${\cal M}_C$ is small and breaks the symmetry of ${\cal M}_A + 
{\cal M}_B$, it is natural to think of its origin in terms of the well-known 
dimension-five operator \cite{w79}
\begin{equation}
{\cal L}_{eff} = {f_{ij} \over 2 \Lambda} (\nu_i \phi^0 - l_i \phi^+)(\nu_j 
\phi^0 - l_j \phi^+) + H.c.,
\end{equation}
where $(\phi^+,\phi^0)$ is the usual Higgs doublet of the Standard Model and 
$\Lambda$ is a very high scale.  As $\phi^0$ picks up a nonzero vacuum 
expectation value $v$, neutrino masses are generated, and if $f_{ij} v^2/
\Lambda = C$ for all $i,j$, ${\cal M}_C$ is obtained.  Since $\Lambda$ is 
presumably of order $10^{16}$ to $10^{18}$ GeV, $C$ is of order $10^{-3}$ 
to $10^{-5}$ eV, and $A-B+3C/2$ is of order 
$10^{-2}$ to 1 eV.  This range of values is just right to encompass all 
three solutions mentioned above.

To justify the assumption that $U_B$ operates in the basis $(\nu_e,\nu_\mu,
\nu_\tau)$, the complete theory of leptons must be discussed.  Under the 
assumed $Z_3$ symmetry, the leptons transform as follows:
\begin{equation}
(\nu,l)_i \to (U_B)_{ij} (\nu,l)_j, ~~~ l^c_k \to l^c_k,
\end{equation}
implemented by 3 Higgs doublets and 1 Higgs triplet:
\begin{equation}
(\phi^0,\phi^-)_i \to (U_B)_{ij} (\phi^0,\phi^-)_j, ~~~  (\xi^{++},\xi^+,\xi^0)
\to (\xi^{++},\xi^+,\xi^0).
\end{equation}
The Yukawa interactions of this model are then given by
\begin{eqnarray}
{\cal L}_Y &=& h_{ij} [\xi^0 \nu_i \nu_j - \xi^+(\nu_i l_j + l_i \nu_j)/
\sqrt 2 + \xi^{++} l_i l_j] \nonumber \\ &+& f_{ij}^k (l_i \phi_j^0 - 
\nu_i \phi_j^-) l^c_k + H.c.
\end{eqnarray}
with
\begin{equation}
h = \pmatrix {a-b & 0 & 0 \cr 0 & a & -b \cr 0 & -b & a}, ~~~ {\cal M}_\nu = 
2 h \langle \xi^0 \rangle,
\end{equation}
and
\begin{equation}
f^k = \pmatrix {a_k-b_k & d_k & d_k \cr -d_k & a_k & -b_k \cr -d_k & -b_k 
& a_k}.
\end{equation}
Note that the $d$ terms are absent in $h$ because it has to be symmetric. 
Assume $v_{1,2} << v_3$, and $d_k << b_k << a_k$, then $V_L {\cal M}_l 
{\cal M}_l^\dagger V_L^\dagger =$ diagonal implies that $V_L$ is nearly 
diagonal.  This justifies the original choice of basis for ${\cal M}_\nu$.

Any model of neutrino mixing implies the presence of lepton flavor violation 
at some level.  In this case, $\phi_1^0$ couples dominantly to $e \tau^c$ 
and $\phi_2^0$ to $\mu \tau^c$.  Taking into account also the other couplings, 
the branching fractions for $\mu \to eee$ and $\mu \to e \gamma$ are estimated 
to be of order $10^{-12}$ and $10^{-11}$ respectively for a Higgs mass 
of 100 GeV.  Both are at the level of present experimental upper bounds.

\section{Conclusions}

The correct form of ${\cal M}_\nu$ is now approximately known.  In the 
$(\nu_e,\nu_\mu,\nu_\tau)$ basis, it obeys the discrete symmetry of Eq.~(34). 
Using Eq.~(32), the phenomenologically successful Eq.~(7) is obtained, which 
has 7 possible limits for ${\cal M}_\nu$.

Assuming some additional symmetry, such as $A_4$ or $Z_3$, 
with possible flavor changing radiative corrections, 
complete theories of leptons (and quarks) may be constructed with 
the prediction of specific neutrino mass patterns and other experimentally 
verifiable consequences.

\section*{Acknowledgements}

I thank Josip Trampetic, Silvio Pallua, and the other organizers of Adriatic 
2003 for their great hospitality at Dubrovnik. 
This work was supported in part by the U.~S.~Department of Energy under 
Grant No.~DE-FG03-94ER40837.

\appendix

\section*{Appendix}

It is amusing to note the parallel between the 5 perfect geometric solids 
and the 5 anomaly-free superstring theories in 10 dimensions.  Whereas the 
former are related among themselves by geometric dualities, the latter are 
related by $S,T,U$ dualities:  Type I $\leftrightarrow$ SO(32), Type IIa 
$\leftrightarrow$ E$_8 \times$E$_8$, and Type IIb is self-dual.  Whereas 
the 5 geometric solids may be embedded in a sphere, the 5 superstring 
theories are believed to be different limits of a single underlying $M$ 
Theory.


\begin{thebibliography}{8.}
\addcontentsline{toc}{section}{References}
\bibitem{kamland} K. Eguchi {\it et al.}, KamLAND Collaboration, Phys. Rev. 
Lett. {\bf 90}, 021802 (2003).
\bibitem{sno} Q. R. Ahmad {\it et al.}, SNO Collaboration, Phys. Rev. Lett. 
{\bf 89}, 011301, 011302 (2002).
\bibitem{atm} For a recent review, see for example C. K. Jung, C. McGrew, 
T. Kajita, and T. Mann, Ann. Rev. Nucl. Part. Sci. {\bf 51}, 451 (2001).
\bibitem{reactor} M. Apollonio {\it et al.}, Phys. Lett. B 466, 415 (1999); 
F. Boehm {\it et al.}, Phys. Rev. D 64, 112001 (2001).
\bibitem{ma02} E. Ma, Phys. Rev. {\bf D66}, 117301 (2002).
\bibitem{gl03} W. Grimus and L. Lavoura, hep-ph/0305309.
\bibitem{mrmm} E. Ma and G. Rajasekaran, Phys. Rev. D 64, 113012 (2001); 
E. Ma, Mod. Phys. Lett. A 17, 289; 627 (2002).
\bibitem{bmv} K. S. Babu, E. Ma, and J. W. F. Valle, Phys. Lett. {\bf B552}, 
207 (2003).
\bibitem{plato} E. Ma, Mod. Phys. Lett. {\bf A17}, 2361 (2002).
\bibitem{wmap} D. N. Spergel {\it et al.}, Astrophys. J. Suppl. {\bf 148}, 
175 (2003).
\bibitem{klapdor} H. V. Klapdor-Kleingrothaus {\it et al.}, Mod. Phys. Lett. 
A 16, 2409 (2001).
\bibitem{kmmrj} J. Kubo, A. Mondragon, M. Mondragon, and E. Rodriguez-Jauregui,
Prog. Theor. Phys. {\bf 109}, 795 (2003).
\bibitem{gld4} W. Grimus and L. Lavoura, Phys. Lett. {\bf B572}, 189 (2003).
\bibitem{ma03} E. Ma, Phys. Rev. Lett. {\bf 90}, 221802 (2003).
\bibitem{msv} M. Maltoni, T. Schwetz, and J. W. F. Valle, Phys. Rev. {\bf D67},
0903003 (2003).
\bibitem{mr03} E. Ma and G. Rajasekaran, Phys. Rev. {\bf D68}, 071302(R) 
(2003).
\bibitem{maz303} E. Ma, hep-ph/0308282.
\bibitem{w79} S. Weinberg, Phys. Rev. Lett. {\bf 43}, 1566 (1979).
\end{thebibliography}
\end{document}